\documentclass[aps,twocolumn,amsmath,superscriptaddress,amssymb,showpacs]{revtex4-1}
\usepackage{bm}
\usepackage{color}
\usepackage{graphicx}
\renewcommand{\vec}[1]{\bm{#1}}

\usepackage[unicode=true,colorlinks=true]{hyperref}

\begin{document}

\title{Spin splitting of electron states in lattice-mismatched (110)-oriented quantum wells}
\author{M.O.~Nestoklon}
\author{S.A.~Tarasenko}
\affiliation{Ioffe Institute, 194021 St. Petersburg, Russia}
\author{R.~Benchamekh}
\affiliation{Tyndall National Institute, Lee Maltings, Dyke Parade, Cork, Ireland}
\author{P.~Voisin}
\affiliation{Laboratoire de Photonique et Nanostructures, CNRS \\ and Universit{\'e} Paris-Saclay, Route de Nozay, 91460 Marcoussis, France}
\begin{abstract}
We show that for lattice-mismatched zinc-blende-type (110)-grown quantum wells a significant contribution to 
the zero-magnetic-field spin splitting of electron subbands comes from strain-induced spin-orbit coupling. 
Combining envelope function theory and atomistic tight-binding approach we calculate spin-orbit splitting constants for realistic quantum wells. It is found that the strain due to lattice mismatch in conventional GaAs/AlGaAs structures may noticeably modify the spin splitting while in InGaAs/GaAs structures it plays a major role and may even change the sign of the spin splitting constant.
\end{abstract}

\maketitle

\section{Introduction}

The $\bm k$-linear spin-orbit splitting of electron states in zinc-blende-type quantum wells (QWs)
is usually discussed in terms of the Rashba spin-orbit coupling stemming from structure inversion asymmetry (SIA)~\cite{Vasko1979,Bychkov84,Sherman03b} and the Dresselhaus spin-orbit coupling originating from $\bm k$-cubic terms in the bulk crystal spectrum
(also named BIA contribution)~\cite{Dresselhaus55,Dyakonov86} and interface inversion asymmetry (IIA)~\cite{Ivchenko96,Krebs98,Vervoort99,Nestoklon06,Tarasenko2015}, or their interplay~\cite{Zawadzki04,Cartoixa06,Nestoklon08,Winkler2012,Alexeev13,Volkov13,Volkov14,Belkov08,Tarasenko09,Volkl2011,Poshakinskiy2013}. Here, $\bm k$ is the electron wave vector. Although it is well known that all QW structures are strained in a varying degree depending on the lattice mismatch between the QW and the buffer layer and strain may also give rise to $\bm k$-linear spin-orbit coupling in bulk crystals~\cite{Pikus_OO,Pikus88}, the effect of strain on spin splitting is commonly neglected. This is largely due to the fact that, in the most studied GaAs/AlGaAs heterostructures, the lattice mismatch does not exceed $0.5\%$ and the whole epilayer adopts the GaAs substrate lattice parameter.

Recently it has been demonstrated experimentally that strain may give a significant contribution 
to spin splitting in real lattice-mismatched heterostructures~\cite{Matsuda11}, but there is a lack of
theoretical studies of this effect. The effect of lattice mismatch on spin splitting
is expected to be more pronounced in QWs of any crystallographic orientation other than (001) since 
shear strain occurring in low-symmetry heterostructures directly couples
the conduction-band and valence-band states~\cite{Pikus_OO,Pikus88}.
Here, we combine the envelope function theory and atomistic tight-binding calculations and
prove that this strain-induced effect can be large as to dominate the spin properties of some 
lattice-mismatched (110)-grown structures. The strain-induced spin-orbit coupling is already 
sizeable for a GaAs/AlGaAs QW when the lattice mismatch is supported by the GaAs well.
The calculations performed for a InGaAs-based QW yield that the strain is the major source of
spin-orbit coupling in the conduction subbands. The resulting Dresselhaus constant of the spin-orbit
splitting in InGaAs-based QWs considerably exceeds that in GaAs/AlGaAs structures and, more importantly,
can be of different sign depending on the buffer layer used.

\section{Effective Hamiltonian}

The effective Hamiltonian describing $\bm k$-linear spin splitting of electron states in a (110)-grown QW
may be generally presented as a sum of three contributions~\cite{Nestoklon12}
\begin{equation}\label{eq:Hso}
H_{so} = \beta\sigma_z k_x + \alpha_+(\sigma_x k_y - \sigma_y k_x) + \alpha_-(\sigma_x k_y + \sigma_yk_x) \,,
\end{equation}
where $\sigma_j$ ($j=x,y,z$) are the Pauli matrices, 
$x \parallel [1\bar{1}0]$ and $y \parallel [00\bar{1}]$ are the in-plane axes, and $z \parallel [110]$ is the growth axis.
The first term on the right-hand side of Eq.~\eqref{eq:Hso} is the $\bm k$-linear Dresselhaus term which originates from BIA and IIA. This term will be drastically affected by strain. The second term describes the isotropic Rashba spin-orbit coupling stemming from SIA. The third term emerges if both the atomic structure of the QW interfaces and the QW structure inversion asymmetry are taken into account. Both the second and the third terms vanish in symmetrically
grown QWs.

Atomistic tight-binding calculations carried out recently for (110) GaAs/Al$_{0.7}$Ga$_{0.3}$As QWs~\cite{Nestoklon12} revealed that the values of $\beta$ and $\alpha_+$ are in a good agreement with the results of the envelope function calculations and $\alpha_-$, which is absent in the isotropic Rashba model, is an order of magnitude smaller than $\alpha_+$. We note that misfit strain was ``switched off'' in the calculations of Ref.~\onlinecite{Nestoklon12} since it is rather weak in GaAs/AlGaAs heterostructures.

The deformation of a bulk zinc-blende-type crystal leads to a $\bm k$-linear spin splitting of the electron spectrum~\cite{Pikus88}.
The corresponding bulk Hamiltonian written in the cubic axes $\tilde{x} \parallel [100]$, $\tilde{y} \parallel [010]$, and $\tilde{z}\parallel [001]$ has the form
\begin{equation}
    H_{\rm str} = \frac12 \left( C_3 \, \vec{\sigma} \cdot \vec{\varphi} + C_3' \, \vec{\sigma} \cdot \vec{\psi}  \right),
\end{equation}
where $C_3$ and $C_3'$ are material constants, $\vec{\varphi}$ and $\vec{\psi}$ are the pseudovectors constructed from the
components of the strain tensor $\vec{\varepsilon}$ and the wave vector $\bm k$,
\begin{equation}
\vec{\varphi} = \left[
\begin{array}{c}
\varepsilon_{\tilde{x}\tilde{y}} k_{\tilde{y}} - \varepsilon_{\tilde{x}\tilde{z}} k_{\tilde{z}} \\
\varepsilon_{\tilde{y}\tilde{z}} k_{\tilde{z}} - \varepsilon_{\tilde{x}\tilde{y}} k_{\tilde{x}} \\
\varepsilon_{\tilde{x}\tilde{z}} k_{\tilde{x}} - \varepsilon_{\tilde{y}\tilde{z}} k_{\tilde{y}}
\end{array}
\right] , \;
\vec{\psi} = \left[
\begin{array}{c}
(\varepsilon_{\tilde{y}\tilde{y}}-\varepsilon_{\tilde{z}\tilde{z}}) k_{\tilde{x}} \\
(\varepsilon_{\tilde{z}\tilde{z}}-\varepsilon_{\tilde{x}\tilde{x}}) k_{\tilde{y}} \\
(\varepsilon_{\tilde{x}\tilde{x}}-\varepsilon_{\tilde{y}\tilde{y}}) k_{\tilde{z}} 
\end{array}
\right] .
\end{equation}
In the $(xyz)$ coordinate frame relevant to (110)-grown structures, the scalar products $\vec{\sigma} \cdot \vec{\varphi}$
and $\vec{\sigma} \cdot \vec{\psi}$ assume the form 
\begin{align}
\vec{\sigma} \cdot \vec{\varphi} &= \frac12 (\varepsilon_{zz}-\varepsilon_{xx})(\sigma_x k_z - \sigma_z k_x) \nonumber \\
&- \varepsilon_{xy} (\sigma_y k_x + \sigma_z k_y) - \varepsilon_{yz} (\sigma_x k_y - \sigma_y k_x) \nonumber \,, \\
\vec{\sigma} \cdot \vec{\psi} &= \frac12 (\varepsilon_{xx} - 2\varepsilon_{yy} + \varepsilon_{zz})(\sigma_x k_z + \sigma_z k_x) \nonumber \\ & - \varepsilon_{xz} (\sigma_x k_x - 2 \sigma_y k_y + \sigma_z k_z ) \,.
\end{align}

Lattice mismatch in a (110)-grown structure leads to the emergence of the strain tensor components 
$\varepsilon_{xx}=\varepsilon_{yy}$ and $\varepsilon_{zz}$ while the off-diagonal components (in the QW coordinate frame) do not occur. The strain tensor can be decomposed in two parts: isotropic part which induces only a change of the band positions, effective mass, and the bulk Dresselhaus constant and anisotropic part with the zero trace which gives rise to $\bm k$-linear spin-orbit coupling. Taking into account that electrons are confined in a QW along the $z$ direction, i.e., $\langle k_z \rangle = 0$, one obtains the strain-induced contribution to the spin Hamiltonian
\begin{equation}
    H_{\rm srt}^{\rm QW} = \frac14 \left( C_3 - C_3' \right) (\varepsilon_{xx} - \varepsilon_{zz}) \sigma_z k_x,
\end{equation}
which corresponds to an additional contribution $\beta_{\text{str}}$ to the Dresselhaus constant, see Eq.~\eqref{eq:Hso},
\begin{equation}
    \beta_{\text{str}} =  \frac14 \left( C_3 - C_3' \right) (\varepsilon_{xx} - \varepsilon_{zz}) \,.
\end{equation}

It is interesting to compare $\beta_{\text{str}}$ with the usual Dresselhaus constant $\beta$ for a standard (110)-grown
GaAs/Al$_{0.3}$Ga$_{0.7}$As QW. The deformation constants for GaAs obtained from the ab initio calculations of Ref.~\cite{Chantis08}
are $C_3 \sim 4-8$ eV$\cdot$\AA{} and $C_3' \sim 2$ eV$\cdot$\AA{}. Recent experimental estimations give $C_3 \approx 8.1$ eV$\cdot$\AA{}
and $C_3'$ is negligibly small~\cite{Beck06}. The typical values of $\beta$ are in the range $7$ -- $15$~meV$\cdot$\AA~\cite{Nestoklon12}
while the lattice mismatch is $\epsilon = \varepsilon_{xx}-\varepsilon_{zz} \sim 1 \cdot 10^{-3}$. Assuming that misfit strain is supported by the GaAs well, we conclude that the strain-induced spin-orbit coupling is only few times smaller than the regular BIA term in such AlGaAs heterostructures.

Conversely, in a In$_{0.2}$Ga$_{0.8}$As/GaAs QW, where the strain $\epsilon \sim 0.02$ is typically 20 times larger and the deformation constant $C_3$ is also larger (see below), the strain-induced spin splitting dominates over the other mechanisms.

To conclude this section, we note that in addition to the renormalization of the Dresselhaus term, non-symmetric strain near interfaces
in realistic structures may also produce electric field independent contributions to the $\alpha_+$ and $\alpha_-$ Rashba terms.
 
\section{Tight-binding calculations}\label{sec:TB}

To calculate the electron dispersion in a QW and extract the parameters of spin-orbit coupling we use the well-established $sp^3d^5s^*$ tight-binding method~\cite{Jancu98,Nestoklon12}.  The method is described in a number of papers and will not be repeated here. Instead, we focus in this section on the procedure of strain incorporation into tight binding.

We use the standard crystallographic coordinate system with a cation atom located at the origin and one of its neighboring anions 
located at $(a_0/4,a_0/4,a_0/4)$, with $a_0$ being the lattice constant. Note, that the opposite choice of the coordinate frame leading to the opposite sign of the bulk Dresselhaus constant $\gamma_c$ is also utilized in literature. 

The strain is microscopically calculated in the valence force field (VFF) approximation~\cite{Keating66} which is able to provide reliable results for small and intermediate strains~\cite{Steiger11}. To model realistic structures we set the lateral lattice constant fixed to mimic the lattice-matched growth on a substrate. Then, we keep the lateral periodic boundary conditions fixed and vary the positions of atoms using the conjugate gradient method to minimize the VFF elastic energy. 
After the minimization we obtain the atomic positions for the fully relaxed structure. This allows us to extract a microscopic strain tensor acting on atomic orbitals using an approach similar to that described in Ref.~\onlinecite{Pryor98}, as explained below. 

For each atom we calculate the ``local strain tensor'' based on the positions of its 4 neighboring atoms.
For a cation C surrounded with 4 anions $\text{A}_i$ ($i = 1 ... 4$) located at arbitrary 
positions, the local strain acting on the cation is defined according to the following procedure. 
First, the nominal anion positions $\boldsymbol{r}_{0i}$ are determined from the bond lengths corresponding 
to the $\text{CA}_i$ bulk lattice parameters in the absence of bond bending. After the structure relaxation, this nominal tetrahedron determined by $\boldsymbol{r}_{0i}$ transforms into the actual one given by the real positions of the atoms $\boldsymbol{r}_{i}$. 
The nominal and actual tetrahedrons can be uniquely characterized by three vectors $\boldsymbol{R}_{j}$ and $\boldsymbol{R}_{(0)j}$
($j = 1 ... 3$), respectively. We choose them as follows:
$\boldsymbol{R}_{(0)1} = \boldsymbol{r}_{(0)2} - \boldsymbol{r}_{(0)1}$, 
$\boldsymbol{R}_{(0)2} = \boldsymbol{r}_{(0)4} - \boldsymbol{r}_{(0)3}$, and 
$\boldsymbol{R}_{(0)3} = [\boldsymbol{r}_{(0)4} + \boldsymbol{r}_{(0)3} - \boldsymbol{r}_{(0)2} -\boldsymbol{r}_{(0)1}]/2$.
Then, we calculate the matrix $\bm T$ connecting the nominal and strained sets, $\boldsymbol{R}_j = \bm T \boldsymbol{R}_{0j}$.
The local strain tensor $\bm \varepsilon$ is then defined by the polar decomposition $\bm T= (1 + \bm \varepsilon) \bm R$, 
where $\bm R$ is the orthogonal matrix of rotation. 

One may check that, for a homogeneously strained bulk binary compound, this approach reproduces the classical definition of strain tensor.
However, the approach allows one to generalize the concept of the strain tensor to the atomic scale.

We notice that the tensor $\bm \varepsilon$ does not fully describe the local atomic
configuration: It is uniquely defined by the relative coordinates of four anions
surrounding a given cation (or, vice versa, by the cation relative coordinates
surrounding a given anion) while the change in the cation position does not 
affect $\bm \varepsilon$. To account for the cation position change we additionally introduce 
an internal strain vector $\boldsymbol{u}$ defined as the displacement of the cation from the 
point equidistant from the surrounding anions and scaled to the unstrained interatomic distance.
For homogeneously strained bulk crystal, the strain tensor $\bm \varepsilon$ and the strain vector $\boldsymbol{u}$ 
are proportional to each other and related by the Kleinman parameter~\cite{Keating66}. However, this is not generally the case
for equilibrium atom positions in a structure with different chemical bonds.

The local strain tensor $\bm \varepsilon$ and the local strain vector $\boldsymbol{u}$ are then incorporated into the tight-binding Hamiltonian. The strain contribution to the tight-binding Hamiltonian has three rather distinct parts. The first one is a scaling of the transfer matrix elements due to the change in the bond lengths~\cite{Jancu98}
\begin{equation}\label{eq:tme_strain}
  V_{n_1,n_2;\, ijk} = V_{n_1,n_2;\, ijk}^0
    \left( \frac{d_{n_1,n_2}}{d_{n_1,n_2}^0} \right)^{n_{ijk}},
\end{equation}
here $n_1$ and $n_2$ are the indices denoting two neighboring atoms, $ijk$ encodes the corresponding Slater off-diagonal parameter, 
$V_{n_1,n_2;\, ijk}^0$ is the transfer matrix element in the unstrained bulk binary compound, 
$d_{n_1,n_2}$ and $d_{n_1,n_2}^0$ are the relaxed interatomic distance and the chemical bond length in the corresponding unstrained compound, and $n_{ijk}$ is the power law scaling exponent~\cite{Jancu98}. For calculations here we use a new set of tight-binding parameters listed in Table~\ref{tbl:tb_par}.

\begin{table}
 \begin{tabular*}{\linewidth}{@{\extracolsep{\fill}}l|rrr}
 \hline
 \hline
 &InAs&GaAs&AlAs\\
 \hline
$              a  $& $  6.0580$& $  5.6500$& $  5.6600$ \\
 \hline
$          E_{s}^a$& $ -6.0738$& $ -5.9820$& $ -6.5474$ \\
$        E_{s^*}^a$& $ 17.2502$& $ 19.4477$& $ 18.9475$ \\
$          E_{s}^c$& $  0.2582$& $ -0.3803$& $  0.3883$ \\
$        E_{s^*}^c$& $ 17.2393$& $ 19.4548$& $ 18.9438$ \\
$          E_{p}^a$& $  2.8784$& $  3.3087$& $  2.9314$ \\
$          E_{d}^a$& $ 11.7833$& $ 13.2015$& $ 12.4961$ \\
$          E_{p}^c$& $  5.6829$& $  6.3801$& $  5.7735$ \\
$          E_{d}^c$& $ 11.7991$& $ 13.2055$& $ 12.4992$ \\
 \hline
$         ss\sigma$& $ -1.5096$& $ -1.6874$& $ -1.8436$ \\
$   s^*_as_c\sigma$& $ -2.0155$& $ -2.1059$& $ -1.7884$ \\
$   s_as^*_c\sigma$& $ -1.1496$& $ -1.5212$& $ -1.3059$ \\
$     s^*s^*\sigma$& $ -3.3608$& $ -3.7170$& $ -3.6128$ \\
$     s_ap_c\sigma$& $  2.2807$& $  2.8846$& $  2.5778$ \\
$     s_cp_a\sigma$& $  2.6040$& $  2.8902$& $  2.7962$ \\
$   s^*_ap_c\sigma$& $  1.9930$& $  2.5294$& $  2.1581$ \\
$   s^*_cp_a\sigma$& $  2.0708$& $  2.3883$& $  2.2397$ \\
$     s_ad_c\sigma$& $ -2.8945$& $ -2.8716$& $ -2.5624$ \\
$     s_cd_a\sigma$& $ -2.3175$& $ -2.2801$& $ -2.3841$ \\
$   s^*_ad_c\sigma$& $ -0.6393$& $ -0.6568$& $ -0.8046$ \\
$   s^*_cd_a\sigma$& $ -0.5949$& $ -0.6113$& $ -0.7492$ \\
$         pp\sigma$& $  3.6327$& $  4.4048$& $  4.1971$ \\
$            pp\pi$& $ -0.9522$& $ -1.4471$& $ -1.3146$ \\
$     p_ad_c\sigma$& $ -1.1156$& $ -1.6035$& $ -1.6473$ \\
$     p_cd_a\sigma$& $ -1.3426$& $ -1.6260$& $ -1.7603$ \\
$        p_ad_c\pi$& $  1.2101$& $  1.8423$& $  1.7647$ \\
$        p_cd_a\pi$& $  1.5282$& $  2.1421$& $  2.1100$ \\
$         dd\sigma$& $ -0.8381$& $ -1.0885$& $ -1.2241$ \\
$            dd\pi$& $  1.9105$& $  2.1560$& $  2.1770$ \\
$         dd\delta$& $ -1.3348$& $ -1.8607$& $ -1.7585$ \\
 \hline
$\Delta_a/3       $& $  0.1558$& $  0.1745$& $  0.1721$ \\
$\Delta_c/3       $& $  0.1143$& $  0.0408$& $  0.0072$ \\
 \hline
 \hline
\end{tabular*}
\caption{Tight-binding parameters used in calculations.}\label{tbl:tb_par}
\end{table}

The second contribution, also introduced in Ref.~\onlinecite{Jancu98}, is the shift of on-site energies proportional to
the hydrostatic component of strain,
\begin{equation}\label{eq:hydr_strain}
\delta E_{\beta} = E_{\beta} - E_{\beta}^0 = - \alpha_{\beta} (E_{\beta}^0 - E_{\text{ref}} ) \frac{\operatorname{Tr}\varepsilon}3 \,,
\end{equation} 
where $E_{\beta}^0$ are the on-site energies in the absence of strain (Table~\ref{tbl:tb_par}),  $\alpha_{\beta}$ are the deformation parameters given in Table~\ref{tbl:strain_par}, and the index  $\beta$ enumerates the orbitals. We define the energy shifts with respect to the reference energy $E_{\text{ref}} = E_{s^*}-6E_{\left\langle1,0,0\right\rangle}$, where $E_{\left\langle1,0,0\right\rangle}=\hbar^2(2\pi/a)^2/2m_0$ and $a$ is the lattice constant. The introduction of $E_{\text{ref}}$ in Eq.~\eqref{eq:hydr_strain} allows us to avoid the recalculation of the deformation parameter $\alpha_{\beta}$ for heterostructures with band offsets. 
The present gauge-invariant formulation is, in the linear limit, strictly equivalent to the one used in Refs.~\onlinecite{Jancu98,Jancu07} for the free electron crystal.  
The choice made for the reference energy is motivated by the aim to keep the positions of the $s^*$ orbitals the same as in the free electron approximation. 

\begin{table}
 \begin{tabular*}{\linewidth}{@{\extracolsep{\fill}}l|rrr}
 \hline
 \hline
 &InAs&GaAs&AlAs\\
 \hline
$   \alpha_s^a    $& $  0.5603$& $  0.0000$& $  0.9720$ \\
$   \alpha_p^a    $& $  1.9539$& $  1.6257$& $  1.8880$ \\
$   \alpha_d^a    $& $  1.7005$& $  2.4531$& $  2.0600$ \\
$   \alpha_s^c    $& $  0.5603$& $  0.0000$& $  0.9720$ \\
$   \alpha_p^c    $& $  1.9539$& $  1.6257$& $  1.8880$ \\
$   \alpha_d^c    $& $  1.7005$& $  2.4531$& $  2.0600$ \\
 \hline
$n_{ss\sigma}     $& $  5.4002$& $  4.5619$& $  2.0880$ \\
$n_{sp\sigma}     $& $  4.4014$& $  3.0363$& $  5.7560$ \\
$n_{sd\sigma}     $& $  6.8053$& $  3.1594$& $  4.4720$ \\
$n_{ss^*\sigma}   $& $  5.8401$& $  3.2676$& $  2.8600$ \\
$n_{s^*p\sigma}   $& $  6.8116$& $  6.9229$& $  3.2240$ \\
$n_{pp\sigma}     $& $  6.9787$& $  6.2602$& $  5.1560$ \\
$n_{pp\pi}        $& $  6.0189$& $  7.0824$& $  2.7960$ \\
$n_{pd\sigma}     $& $  2.7559$& $  3.5344$& $  5.5920$ \\
$n_{pd\pi}        $& $  6.0212$& $  7.3976$& $  4.8080$ \\
 \hline
$\pi_{001}        $& $  0.0952$& $  0.1476$& $  0.1000$ \\
$\pi_{111}        $& $  0.1456$& $  0.1588$& $  0.1160$ \\
 \hline
 \hline
 \end{tabular*}
\caption{Deformation tight-binding parameters used in calculations.
Other parameters are $\alpha_{s^*}=2.0$ and $n_{s^*s^*\sigma}=n_{s^*d\sigma}=n_{dd\sigma}=n_{dd\pi}=n_{dd\delta}= 2.0$
}\label{tbl:strain_par}
\end{table}

The third contribution is related to the strain-induced splitting of the on-site energies of the degenerate orbitals $p$ and $d$~\cite{Jancu07,Boykin10,Niquet09,Zielinski12}. The contribution has not been analyzed in detail so far. In Ref.~\onlinecite{Jancu07}, a simplified approach has been considered: The splittings were assumed to be proportional to the strain tensor. Here, we generalize this approach by introducing the corrections proportional to the local strain tensor $\bm \varepsilon$ and the local strain vector $\vec{u}$.

Using the method of invariants one can show that the corresponding contribution to
the $p$-orbital same-atom block in the tight-binding Hamiltonian in the basis of the
functions $p_x$, $p_y$, and $p_z$ has the form
\begin{equation}\label{ham_str_n_int}
\delta H=
\left(\begin{array}{ccc}
\lambda_1(\sqrt3\varepsilon_1-\varepsilon_2)
   &\lambda_2\varepsilon_{xy} + \xi u_z
       &\lambda_2\varepsilon_{zx} + \xi u_y\\
\lambda_2\varepsilon_{xy}+\xi u_z
   &-\lambda_1(\sqrt3\varepsilon_1+\varepsilon_2)
       &\lambda_2\varepsilon_{yz} +\xi u_x\\
\lambda_2\varepsilon_{zx}+\xi u_y
   &\lambda_2\varepsilon_{yz}+\xi u_x
       &2\lambda_1\varepsilon_2
\end{array}\right),
\end{equation}
where $\varepsilon_1 = \sqrt{3}(\varepsilon_{xx}-\varepsilon_{yy})$, $\varepsilon_2=2\varepsilon_{zz}-\varepsilon_{xx}-\varepsilon_{yy}$,
and $\lambda_j$ ($j=1,2$) are parameters, and we assume $\xi = \pm\lambda_{2}$ for anions and cations, respectively. To make the parametrization space more compact we assume that the parameters $\lambda_j$ for anions and cations are connected to each other by
\begin{align}
  \lambda_{1}^a & =  \frac12 (E_{p}^a-E_{\text{ref}}^a) \pi_{100} \,,\; 
	\lambda_{1}^c  =  \frac12 (E_{p}^c-E_{\text{ref}}^c) \pi_{100} \,, \\
  \lambda_{2}^a & = -\frac83 (E_{p}^a-E_{\text{ref}}^a) \pi_{111} \,,\;
	\lambda_{2}^c  = -\frac83 (E_{p}^c-E_{\text{ref}}^c) \pi_{111} \,. \nonumber
\end{align}
The deformation parameters $\pi_{100}$ and $\pi_{111}$ for several binary compounds are listed in Table~\ref{tbl:strain_par}.
We note that similar splitting occurs for the $d$ orbitals as well. However, to fit the conduction-band and valence-band deformation-potential constants at the $\Gamma$ point of the Brillouin zone it is sufficient to consider the splitting of the $p$ orbitals.
Therefore, to simplify calculations we neglect the splitting of the $d$ orbitals. We keep in mind that for fitting simultaneously the deformation-potential constants at the $\Gamma$, $X$ and $L$ points a complete set of parameters should be used.

The determination of the strain-related parameters of the tight-binding Hamiltonian is a challenging task because of the small number of available well documented deformation-potential constants. In fact, the deformation-potential constants at the Brillouin zone center do not provide enough information to uniquely determine all the tight-binding parameters. We expect that any parametrization providing the correct values of the conduction-band deformation-potential constant $a_c$ and the valence-band deformation-potential constants
$a_v$, $b$ and $d$ yields satisfactory results the strain-induced spin-orbit coupling. Therefore, we adopt the approach 
described in Ref.~\onlinecite{Raouafi16} and numerically fit the strain-related tight-binding parameters to to reproduce the recommended values of the deformation-potential constants in the Brillouin zone center given in Ref.~\onlinecite{Meyer}.
The obtain parameters are presented in Table~\ref{tbl:strain_par}. Table~\ref{tbl:kp_par} summarize the band gaps $E_g$, the effective electron masses at the $\Gamma$ point $m_e$, the Dresselhaus constants $\gamma_c$, and the deformation-potential constants $a_c$, $a_v$, $b$, $d$, and $C_3$ for some bulk binary and ternary compounds obtained from the tight-binding calculations with the parameters listed in Table~\ref{tbl:tb_par} and Table~\ref{tbl:strain_par}.

For alloys, the tight-binding needs a special care to be taken to reproduce the band gap bowing properly~\cite{Shim98}. 
Here, we use an original interpolation scheme \cite{Nestoklon16} to construct the alloy A$_x$B$_{1-x}$C tight-binding parameters
in the virtual crystal approximation from the tight-binding parameters of the binary compounds AC and BC.
First, the lattice constant of the alloy is found as the linear interpolation between the binaries.
Then, we calculate the strain contributions as described above and construct the parameters of the AC and BC
materials strained to the lattice constant of the alloy. Finally, the tight-binding parameters of the alloy are determined 
as the linear interpolations of the parameters of the strained binary materials. We note that, unlike the approach 
from Ref.~\onlinecite{Shim98}, this procedure provides the correct bowing without the introduction of additional parameters.

\begin{table}
\begin{tabular}{c|rrrrr}
\hline\hline
                 &       GaAs &       AlAs &   InAs & Al$_{0.3}$Ga$_{0.7}$As & In$_{0.2}$Ga$_{0.8}$As \\
\hline
$E_g$        & $1.519$  & $3.130$ & $0.417$ & $2.000$&  $1.207$\\
$m_e$        & $0.0665$ & $0.158$ & $0.0235$& $0.0904$& $0.0519$\\
$-\gamma_c$   & $24.21 $ & $9.12 $ & $45.39$ & $16.04 $& $28.61$ \\
\hline                                                
$a_c$        & $-7.17$  & $-5.64$ & $-5.08$ & $-6.74$ & $-6.62$\\
$a_v$        & $ 1.60$  & $ 2.47$ & $ 1.00$ & $ 1.63$ & $ 1.22$   \\
$b$          & $-2.00$  & $-2.30$ & $-1.80$ & $-0.74$ & $-1.35$ \\
$d$          & $-4.80$  & $-3.40$ & $-3.60$ & $-4.39$ & $-4.53$ \\
$C_3$        & $ 8.12$  & $-3.34$ & $104.5$ & $-2.65$ & $15.10$ \\
\hline\hline
\end{tabular}
\caption{The band gaps $E_g$ (in eV), the effective electron masses at the $\Gamma$ point $m_e$ (in the units of free electron mass), the bulk Dresselhaus constants $\gamma_c$ (in eV$\cdot$\AA$^3$) \cite{Note1}, the deformation-potential constants $a_c$, $a_v$, $b$ and $d$ (in eV), and
$C_3$ (in eV$\cdot$\AA) computed using the tight-binding parameters from Table~\ref{tbl:tb_par} and Table~\ref{tbl:strain_par}. 
}\label{tbl:kp_par}
\end{table}

We also note that the standard $sp^3d^5s^*$ tight-binding model~\cite{Jancu98} does not take into
account the spin-orbit coupling of the $p$ and $d$ orbitals which yields the major contribution
to the $C_3'$ constant~\cite{Pikus88}. The missing of $C_3'$ has the same origin 
as the missing of the $\bm k$-linear spin splitting of the $\Gamma_8$ valence band in the bulk 
crystal~\cite{Pikus88,Boykin98}. Possible solution of this problem proposed by 
Boykin~\cite{Boykin98} is based on the consideration of second-nearest neighboring atoms 
and has no straightforward extension to strained heterostructures.

\section{Results}

We use the procedure described in Ref.~\onlinecite{Nestoklon12} to extract the constants 
of the spin-orbit Hamiltonian \eqref{eq:Hso} from the tight-binding calculations for
GaAs-based and InGaAs-based QW structures. As distinct from previous calculations, 
we now include the atomistic strain as described in Section~\ref{sec:TB}. The results
show that the constants $\alpha_{\pm}$ related to structure inversion asymmetry are almost 
independent of the strain present in the QW. Therefore, we focus below on the 
Dresselhaus parameter $\beta$ and consider symmetric QWs.

\subsection{GaAs/AlGaAs quantum wells}

\begin{figure}
\includegraphics[width=\linewidth]{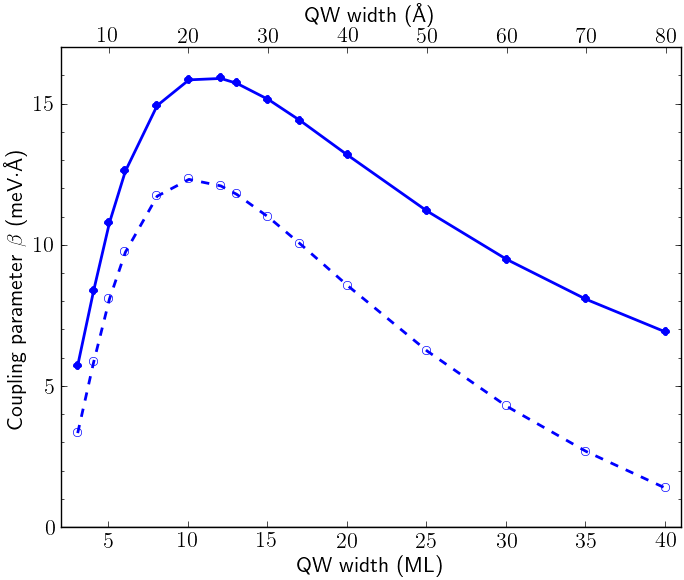}%
\caption{Dresselhaus constant $\beta$ as a function of the QW thickness (in monolayers, ML) calculated for 
Al$_{0.3}$Ga$_{0.7}$As/GaAs/Al$_{0.3}$Ga$_{0.7}$As QWs with different strain configurations.
Solid line shows the results for QW structures with the lattice constant corresponding to GaAs, 
$a_0=5.65$~\AA. Dashed line shows the results for QWs lattice-matched to AlAs with the lattice constant 
$a_0=5.66$.
}
\label{fig:AlGaAs}
\end{figure}

The Dresselhaus constant $\beta$ as a function of the QW thickness determined for
AlGaAs/GaAs/AlGaAs QWs with different strain configurations is shown in Fig.~\ref{fig:AlGaAs}.
Here, we present the results for the same structures but lattice-matched either to GaAs 
(solid line) or to AlAs (dashed line)~\footnote{The coordinate system is changed with respect to the
one used in Ref.~\cite{Nestoklon12} which results in the opposite sign of the Dresselhaus constant.}.

From Fig.~\ref{fig:AlGaAs} one may conclude that the widely adopted consideration of GaAs/AlGaAs heterostructures as unstrained systems  is not completely satisfied for the analysis of spin splitting. The ratio of the Dresselhaus constants $\beta$ for the structures lattice-matched to GaAs and AlAs would exceed a factor 2 for a 50-\AA-wide QW while the strain itself is only $1\cdot10^{-3}$. 
Note that the observed dependence of $\beta$ on the QW thickness actually reflects the redistribution of electron presence probability between the well and the barriers.

\subsection{InGaAs/GaAs quantum wells}

\begin{figure}
\includegraphics[width=\linewidth]{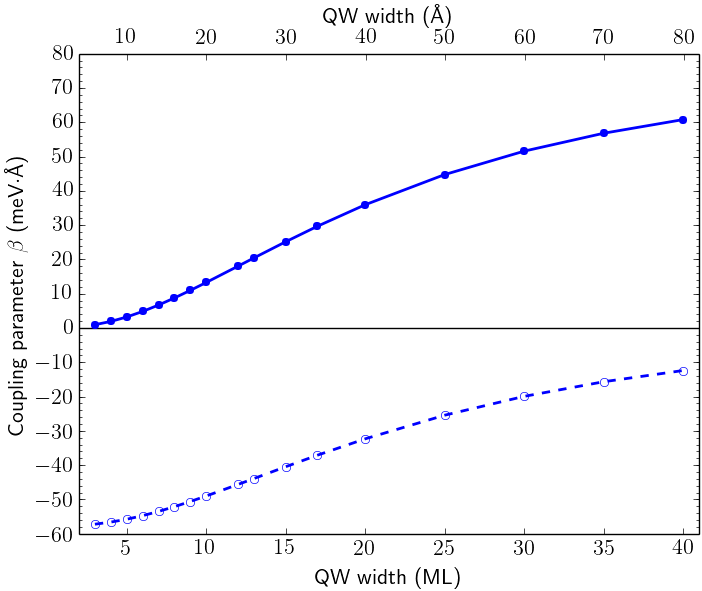}%
\caption{Dresselhaus constant $\beta$ as a function of the QW thickness calculated for 
GaAs/In$_{0.2}$Ga$_{0.8}$As/GaAs QWs grown on GaAs buffer layer (solid lines) and 
In$_{0.2}$Ga$_{0.8}$As buffer layer (dashed lines).}\label{fig:InGaAs}
\end{figure} 

While the strain contribution to the spin splitting in GaAs/AlGaAs QWs is still a correction,
it is natural to expect that in heterostructures grown from compounds with a significant lattice mismatch,
like InGaAs/GaAs QWs, the strain contribution dominates the spin splitting. 

In addition to larger lattice mismatch, the deformation constant $C_3$ in InAs is about an order of magnitude larger than that in GaAs,
see Table~\ref{tbl:kp_par}. The large value of $C_3$ can be explained from the $\bm k$$\cdot$$\bm p$ perturbation theory 
for bulk crystals where the major contribution to $C_3$ is given by~\cite{Pikus88}
\begin{equation}\label{eq:C3}
 C_3 = \frac43 \frac{C_2 P \Delta}{E_g(E_g+\Delta)} \,,
\end{equation}
where $C_2$ is the interband deformation-potential constant, $P$ is the Kane matrix element, and $\Delta$ is the spin-orbit splitting of the valence band. The growth of $C_3$ for InAs as compared to that for GaAs is caused by the decrease in $E_g$ and the increase in $C_2$.  

To elaborate this expectation we calculate the Dresselhaus constant $\beta$ for GaAs/In$_{0.2}$Ga$_{0.8}$As/GaAs QW
structures lattice-matched to GaAs and In$_{0.2}$Ga$_{0.8}$As. The dependence of the Dresselhaus constants on the QW thickness
for both strain configurations are shown in Fig.~\ref{fig:InGaAs}. The range of spin splittings is significantly larger than
that in GaAs/AlGaAs structures and the Dresselhaus constant has a different dependence on the QW thickness.
Importantly, the sign of $\beta$ is opposite for the QW structures lattice-matched to GaAs and InGaAs layers. In the first case, $\beta$ tends to zero for narrow wells and saturates to the constant $\beta_{\rm srt}$ of the strained bulk InGaAs layer for wide wells. In the latter case, the behavior is opposite: $\beta$ tends to the constant $\beta_{\rm srt}$ of the strained bulk GaAs layer for narrow wells and vanishes for wide wells. The effect of quantum confinement given by $\langle k_z^2 \rangle$, which is important for GaAlAs QWs, 
is masked by the interplay between the strain and the electron probability of presence in the well and the barriers. Actually, the expected positive value of $\beta$ for thick unstrained wells (dashed line in Fig.~\ref{fig:InGaAs}) is recovered only for very large thicknesses ($>60$ ML).

\section{Conclusion}

In conclusion, we have performed atomistic calculations of the spin-orbit splitting of electron subbands in III-V (110)-grown quantum wells and revealed the important role of strain which naturally occurs in heterostructures. The strain contribution to the spin-orbit 
coupling noticeably renormalizes the Dresselhaus constant in GaAs/AlGaAs QWs, which are commonly treated as nearly unstrained, and dominates the spin splitting in InGaAs/GaAs QWs with a rather large lattice constant mismatch. Strain engineering thus opens a way to control the spin splittings in two-dimensional electron gas in semiconductor heterostructures.

\paragraph*{Acknowledgments.}  This work was partly supported by the Russian-French International Laboratory ILNACS and the
 RFBR (projects 14-02-00123 and 15-32-20828).

\bibliography{SIA_110_strain}

\begin{thebibliography}{41}%
\makeatletter
\providecommand \@ifxundefined [1]{%
 \@ifx{#1\undefined}
}%
\providecommand \@ifnum [1]{%
 \ifnum #1\expandafter \@firstoftwo
 \else \expandafter \@secondoftwo
 \fi
}%
\providecommand \@ifx [1]{%
 \ifx #1\expandafter \@firstoftwo
 \else \expandafter \@secondoftwo
 \fi
}%
\providecommand \natexlab [1]{#1}%
\providecommand \enquote  [1]{``#1''}%
\providecommand \bibnamefont  [1]{#1}%
\providecommand \bibfnamefont [1]{#1}%
\providecommand \citenamefont [1]{#1}%
\providecommand \href@noop [0]{\@secondoftwo}%
\providecommand \href [0]{\begingroup \@sanitize@url \@href}%
\providecommand \@href[1]{\@@startlink{#1}\@@href}%
\providecommand \@@href[1]{\endgroup#1\@@endlink}%
\providecommand \@sanitize@url [0]{\catcode `\\12\catcode `\$12\catcode
  `\&12\catcode `\#12\catcode `\^12\catcode `\_12\catcode `\%12\relax}%
\providecommand \@@startlink[1]{}%
\providecommand \@@endlink[0]{}%
\providecommand \url  [0]{\begingroup\@sanitize@url \@url }%
\providecommand \@url [1]{\endgroup\@href {#1}{\urlprefix }}%
\providecommand \urlprefix  [0]{URL }%
\providecommand \Eprint [0]{\href }%
\providecommand \doibase [0]{http://dx.doi.org/}%
\providecommand \selectlanguage [0]{\@gobble}%
\providecommand \bibinfo  [0]{\@secondoftwo}%
\providecommand \bibfield  [0]{\@secondoftwo}%
\providecommand \translation [1]{[#1]}%
\providecommand \BibitemOpen [0]{}%
\providecommand \bibitemStop [0]{}%
\providecommand \bibitemNoStop [0]{.\EOS\space}%
\providecommand \EOS [0]{\spacefactor3000\relax}%
\providecommand \BibitemShut  [1]{\csname bibitem#1\endcsname}%
\let\auto@bib@innerbib\@empty
\bibitem [{\citenamefont {Vasko}\ and\ \citenamefont
  {Prima}(1979)}]{Vasko1979}%
  \BibitemOpen
  \bibfield  {author} {\bibinfo {author} {\bibfnamefont {F.~T.}\ \bibnamefont
  {Vasko}}\ and\ \bibinfo {author} {\bibfnamefont {N.~A.}\ \bibnamefont
  {Prima}},\ }\href@noop {} {\bibfield  {journal} {\bibinfo  {journal} {Fiz.
  Tverd. Tela}\ }\textbf {\bibinfo {volume} {21}},\ \bibinfo {pages} {1734}
  (\bibinfo {year} {1979})},\ \translation{Sov. Phys. Solid State {\bf 21}, 994
  (1979)}\BibitemShut {NoStop}%
\bibitem [{\citenamefont {Bychkov}\ and\ \citenamefont
  {Rashba}(1984)}]{Bychkov84}%
  \BibitemOpen
  \bibfield  {author} {\bibinfo {author} {\bibfnamefont {Y.~A.}\ \bibnamefont
  {Bychkov}}\ and\ \bibinfo {author} {\bibfnamefont {E.}~\bibnamefont
  {Rashba}},\ }\href@noop {} {\bibfield  {journal} {\bibinfo  {journal} {Pis'ma
  Zh. Eksp. Teor. Fiz.}\ }\textbf {\bibinfo {volume} {39}},\ \bibinfo {pages}
  {66} (\bibinfo {year} {1984})},\ \translation{JETP Lett. {\bf 39}, 78
  (1984)}\BibitemShut {NoStop}%
\bibitem [{\citenamefont {Sherman}(2003)}]{Sherman03b}%
  \BibitemOpen
  \bibfield  {author} {\bibinfo {author} {\bibfnamefont {E.~Y.}\ \bibnamefont
  {Sherman}},\ }\href {\doibase http://dx.doi.org/10.1063/1.1533839} {\bibfield
   {journal} {\bibinfo  {journal} {Applied Physics Letters}\ }\textbf {\bibinfo
  {volume} {82}},\ \bibinfo {pages} {209} (\bibinfo {year} {2003})}\BibitemShut
  {NoStop}%
\bibitem [{\citenamefont {Dresselhaus}(1955)}]{Dresselhaus55}%
  \BibitemOpen
  \bibfield  {author} {\bibinfo {author} {\bibfnamefont {G.}~\bibnamefont
  {Dresselhaus}},\ }\href {\doibase 10.1103/PhysRev.100.580} {\bibfield
  {journal} {\bibinfo  {journal} {Phys. Rev.}\ }\textbf {\bibinfo {volume}
  {100}},\ \bibinfo {pages} {580} (\bibinfo {year} {1955})}\BibitemShut
  {NoStop}%
\bibitem [{\citenamefont {D'yakonov}\ and\ \citenamefont
  {Kachorovskii}(1986)}]{Dyakonov86}%
  \BibitemOpen
  \bibfield  {author} {\bibinfo {author} {\bibfnamefont {M.~I.}\ \bibnamefont
  {D'yakonov}}\ and\ \bibinfo {author} {\bibfnamefont {V.~Y.}\ \bibnamefont
  {Kachorovskii}},\ }\href@noop {} {\bibfield  {journal} {\bibinfo  {journal}
  {Fiz. Tekh. Poluprovodn.}\ }\textbf {\bibinfo {volume} {20}},\ \bibinfo
  {pages} {178} (\bibinfo {year} {1986})},\ \translation{Sov. Phys. Semicond.
  {\bf 20}, 110 (1986)}\BibitemShut {NoStop}%
\bibitem [{\citenamefont {Ivchenko}\ \emph {et~al.}(1996)\citenamefont
  {Ivchenko}, \citenamefont {Kaminski},\ and\ \citenamefont
  {R\"ossler}}]{Ivchenko96}%
  \BibitemOpen
  \bibfield  {author} {\bibinfo {author} {\bibfnamefont {E.~L.}\ \bibnamefont
  {Ivchenko}}, \bibinfo {author} {\bibfnamefont {A.~Y.}\ \bibnamefont
  {Kaminski}}, \ and\ \bibinfo {author} {\bibfnamefont {U.}~\bibnamefont
  {R\"ossler}},\ }\href {\doibase 10.1103/PhysRevB.54.5852} {\bibfield
  {journal} {\bibinfo  {journal} {Phys. Rev. B}\ }\textbf {\bibinfo {volume}
  {54}},\ \bibinfo {pages} {5852} (\bibinfo {year} {1996})}\BibitemShut
  {NoStop}%
\bibitem [{\citenamefont {Krebs}\ \emph {et~al.}(1998)\citenamefont {Krebs},
  \citenamefont {Rondi}, \citenamefont {Gentner}, \citenamefont {Goldstein},\
  and\ \citenamefont {Voisin}}]{Krebs98}%
  \BibitemOpen
  \bibfield  {author} {\bibinfo {author} {\bibfnamefont {O.}~\bibnamefont
  {Krebs}}, \bibinfo {author} {\bibfnamefont {D.}~\bibnamefont {Rondi}},
  \bibinfo {author} {\bibfnamefont {J.~L.}\ \bibnamefont {Gentner}}, \bibinfo
  {author} {\bibfnamefont {L.}~\bibnamefont {Goldstein}}, \ and\ \bibinfo
  {author} {\bibfnamefont {P.}~\bibnamefont {Voisin}},\ }\href {\doibase
  10.1103/PhysRevLett.80.5770} {\bibfield  {journal} {\bibinfo  {journal}
  {Phys. Rev. Lett.}\ }\textbf {\bibinfo {volume} {80}},\ \bibinfo {pages}
  {5770} (\bibinfo {year} {1998})}\BibitemShut {NoStop}%
\bibitem [{\citenamefont {Vervoort}\ \emph {et~al.}(1999)\citenamefont
  {Vervoort}, \citenamefont {Ferreira},\ and\ \citenamefont
  {Voisin}}]{Vervoort99}%
  \BibitemOpen
  \bibfield  {author} {\bibinfo {author} {\bibfnamefont {L.}~\bibnamefont
  {Vervoort}}, \bibinfo {author} {\bibfnamefont {R.}~\bibnamefont {Ferreira}},
  \ and\ \bibinfo {author} {\bibfnamefont {P.}~\bibnamefont {Voisin}},\ }\href
  {http://stacks.iop.org/0268-1242/14/i=3/a=004} {\bibfield  {journal}
  {\bibinfo  {journal} {Semiconductor Science and Technology}\ }\textbf
  {\bibinfo {volume} {14}},\ \bibinfo {pages} {227} (\bibinfo {year}
  {1999})}\BibitemShut {NoStop}%
\bibitem [{\citenamefont {Nestoklon}\ \emph {et~al.}(2006)\citenamefont
  {Nestoklon}, \citenamefont {Golub},\ and\ \citenamefont
  {Ivchenko}}]{Nestoklon06}%
  \BibitemOpen
  \bibfield  {author} {\bibinfo {author} {\bibfnamefont {M.~O.}\ \bibnamefont
  {Nestoklon}}, \bibinfo {author} {\bibfnamefont {L.~E.}\ \bibnamefont
  {Golub}}, \ and\ \bibinfo {author} {\bibfnamefont {E.~L.}\ \bibnamefont
  {Ivchenko}},\ }\href {\doibase 10.1103/PhysRevB.73.235334} {\bibfield
  {journal} {\bibinfo  {journal} {Phys. Rev. B}\ }\textbf {\bibinfo {volume}
  {73}},\ \bibinfo {eid} {235334} (\bibinfo {year} {2006})}\BibitemShut
  {NoStop}%
\bibitem [{\citenamefont {Tarasenko}\ \emph {et~al.}(2015)\citenamefont
  {Tarasenko}, \citenamefont {Durnev}, \citenamefont {Nestoklon}, \citenamefont
  {Ivchenko}, \citenamefont {Luo},\ and\ \citenamefont
  {Zunger}}]{Tarasenko2015}%
  \BibitemOpen
  \bibfield  {author} {\bibinfo {author} {\bibfnamefont {S.~A.}\ \bibnamefont
  {Tarasenko}}, \bibinfo {author} {\bibfnamefont {M.~V.}\ \bibnamefont
  {Durnev}}, \bibinfo {author} {\bibfnamefont {M.~O.}\ \bibnamefont
  {Nestoklon}}, \bibinfo {author} {\bibfnamefont {E.~L.}\ \bibnamefont
  {Ivchenko}}, \bibinfo {author} {\bibfnamefont {J.-W.}\ \bibnamefont {Luo}}, \
  and\ \bibinfo {author} {\bibfnamefont {A.}~\bibnamefont {Zunger}},\ }\href
  {\doibase 10.1103/PhysRevB.91.081302} {\bibfield  {journal} {\bibinfo
  {journal} {Phys. Rev. B}\ }\textbf {\bibinfo {volume} {91}},\ \bibinfo
  {pages} {081302} (\bibinfo {year} {2015})}\BibitemShut {NoStop}%
\bibitem [{\citenamefont {Zawadzki}\ and\ \citenamefont
  {Pfeffer}(2004)}]{Zawadzki04}%
  \BibitemOpen
  \bibfield  {author} {\bibinfo {author} {\bibfnamefont {W.}~\bibnamefont
  {Zawadzki}}\ and\ \bibinfo {author} {\bibfnamefont {P.}~\bibnamefont
  {Pfeffer}},\ }\href
  {http://iopscience.iop.org/article/10.1088/0268-1242/19/1/R01} {\bibfield
  {journal} {\bibinfo  {journal} {Semicond. Sci. Technol.}\ }\textbf {\bibinfo
  {volume} {19}},\ \bibinfo {pages} {R1} (\bibinfo {year} {2004})}\BibitemShut
  {NoStop}%
\bibitem [{\citenamefont {Cartoix\`{a}}\ \emph {et~al.}(2006)\citenamefont
  {Cartoix\`{a}}, \citenamefont {Wang}, \citenamefont {Ting},\ and\
  \citenamefont {Chang}}]{Cartoixa06}%
  \BibitemOpen
  \bibfield  {author} {\bibinfo {author} {\bibfnamefont {X.}~\bibnamefont
  {Cartoix\`{a}}}, \bibinfo {author} {\bibfnamefont {L.-W.}\ \bibnamefont
  {Wang}}, \bibinfo {author} {\bibfnamefont {D.-Y.}\ \bibnamefont {Ting}}, \
  and\ \bibinfo {author} {\bibfnamefont {Y.-C.}\ \bibnamefont {Chang}},\ }\href
  {http://journals.aps.org/prb/abstract/10.1103/PhysRevB.73.205341} {\bibfield
  {journal} {\bibinfo  {journal} {Phys. Rev. B}\ }\textbf {\bibinfo {volume}
  {73}},\ \bibinfo {pages} {205341} (\bibinfo {year} {2006})}\BibitemShut
  {NoStop}%
\bibitem [{\citenamefont {Nestoklon}\ \emph {et~al.}(2008)\citenamefont
  {Nestoklon}, \citenamefont {Ivchenko}, \citenamefont {Jancu},\ and\
  \citenamefont {Voisin}}]{Nestoklon08}%
  \BibitemOpen
  \bibfield  {author} {\bibinfo {author} {\bibfnamefont {M.~O.}\ \bibnamefont
  {Nestoklon}}, \bibinfo {author} {\bibfnamefont {E.~L.}\ \bibnamefont
  {Ivchenko}}, \bibinfo {author} {\bibfnamefont {J.-M.}\ \bibnamefont {Jancu}},
  \ and\ \bibinfo {author} {\bibfnamefont {P.}~\bibnamefont {Voisin}},\ }\href
  {http://journals.aps.org/prb/abstract/10.1103/PhysRevB.77.155328} {\bibfield
  {journal} {\bibinfo  {journal} {Phys. Rev. B}\ }\textbf {\bibinfo {volume}
  {77}},\ \bibinfo {pages} {155328} (\bibinfo {year} {2008})}\BibitemShut
  {NoStop}%
\bibitem [{\citenamefont {Winkler}\ \emph {et~al.}(2012)\citenamefont
  {Winkler}, \citenamefont {Wang}, \citenamefont {Lin},\ and\ \citenamefont
  {Chu}}]{Winkler2012}%
  \BibitemOpen
  \bibfield  {author} {\bibinfo {author} {\bibfnamefont {R.}~\bibnamefont
  {Winkler}}, \bibinfo {author} {\bibfnamefont {L.}~\bibnamefont {Wang}},
  \bibinfo {author} {\bibfnamefont {Y.}~\bibnamefont {Lin}}, \ and\ \bibinfo
  {author} {\bibfnamefont {C.}~\bibnamefont {Chu}},\ }\href {\doibase
  http://dx.doi.org/10.1016/j.ssc.2012.09.002} {\bibfield  {journal} {\bibinfo
  {journal} {Solid State Commun.}\ }\textbf {\bibinfo {volume} {152}},\
  \bibinfo {pages} {2096 } (\bibinfo {year} {2012})}\BibitemShut {NoStop}%
\bibitem [{\citenamefont {Alexeev}(2013)}]{Alexeev13}%
  \BibitemOpen
  \bibfield  {author} {\bibinfo {author} {\bibfnamefont {P.}~\bibnamefont
  {Alexeev}},\ }\href@noop {} {\bibfield  {journal} {\bibinfo  {journal}
  {Pis’ma Zh. Eksp. Teor. Fiz.}\ }\textbf {\bibinfo {volume} {98}},\ \bibinfo
  {pages} {84} (\bibinfo {year} {2013})},\ \translation{JETP Lett. {\bf 98}, 92
  (2013)}\BibitemShut {NoStop}%
\bibitem [{\citenamefont {Devizorova}\ and\ \citenamefont
  {Volkov}(2013)}]{Volkov13}%
  \BibitemOpen
  \bibfield  {author} {\bibinfo {author} {\bibfnamefont {Z.}~\bibnamefont
  {Devizorova}}\ and\ \bibinfo {author} {\bibfnamefont {V.}~\bibnamefont
  {Volkov}},\ }\href@noop {} {\bibfield  {journal} {\bibinfo  {journal}
  {Pis’ma Zh. Eksp. Teor. Fiz.}\ }\textbf {\bibinfo {volume} {98}},\ \bibinfo
  {pages} {101} (\bibinfo {year} {2013})},\ \translation{JETP Lett. {\bf 98},
  110 (2013)}\BibitemShut {NoStop}%
\bibitem [{\citenamefont {Devizorova}\ \emph {et~al.}(2014)\citenamefont
  {Devizorova}, \citenamefont {Shchepetilnikov}, \citenamefont {Nefyodov},
  \citenamefont {Volkov},\ and\ \citenamefont {Kukushkin}}]{Volkov14}%
  \BibitemOpen
  \bibfield  {author} {\bibinfo {author} {\bibfnamefont {Z.}~\bibnamefont
  {Devizorova}}, \bibinfo {author} {\bibfnamefont {A.}~\bibnamefont
  {Shchepetilnikov}}, \bibinfo {author} {\bibfnamefont {Y.}~\bibnamefont
  {Nefyodov}}, \bibinfo {author} {\bibfnamefont {V.}~\bibnamefont {Volkov}}, \
  and\ \bibinfo {author} {\bibfnamefont {I.~V.}\ \bibnamefont {Kukushkin}},\
  }\href@noop {} {\bibfield  {journal} {\bibinfo  {journal} {Pis’ma Zh. Eksp.
  Teor. Fiz.}\ }\textbf {\bibinfo {volume} {100}},\ \bibinfo {pages} {102}
  (\bibinfo {year} {2014})},\ \translation{JETP Lett. {\bf 100}, 111
  (2014)}\BibitemShut {NoStop}%
\bibitem [{\citenamefont {Bel'kov}\ \emph {et~al.}(2008)\citenamefont
  {Bel'kov}, \citenamefont {Olbrich}, \citenamefont {Tarasenko}, \citenamefont
  {Schuh}, \citenamefont {Wegscheider}, \citenamefont {Korn}, \citenamefont
  {Sch\"uller}, \citenamefont {Weiss}, \citenamefont {Prettl},\ and\
  \citenamefont {Ganichev}}]{Belkov08}%
  \BibitemOpen
  \bibfield  {author} {\bibinfo {author} {\bibfnamefont {V.~V.}\ \bibnamefont
  {Bel'kov}}, \bibinfo {author} {\bibfnamefont {P.}~\bibnamefont {Olbrich}},
  \bibinfo {author} {\bibfnamefont {S.~A.}\ \bibnamefont {Tarasenko}}, \bibinfo
  {author} {\bibfnamefont {D.}~\bibnamefont {Schuh}}, \bibinfo {author}
  {\bibfnamefont {W.}~\bibnamefont {Wegscheider}}, \bibinfo {author}
  {\bibfnamefont {T.}~\bibnamefont {Korn}}, \bibinfo {author} {\bibfnamefont
  {C.}~\bibnamefont {Sch\"uller}}, \bibinfo {author} {\bibfnamefont
  {D.}~\bibnamefont {Weiss}}, \bibinfo {author} {\bibfnamefont
  {W.}~\bibnamefont {Prettl}}, \ and\ \bibinfo {author} {\bibfnamefont {S.~D.}\
  \bibnamefont {Ganichev}},\ }\href
  {http://journals.aps.org/prl/abstract/10.1103/PhysRevLett.100.176806}
  {\bibfield  {journal} {\bibinfo  {journal} {Phys. Rev. Lett.}\ }\textbf
  {\bibinfo {volume} {100}},\ \bibinfo {pages} {176806} (\bibinfo {year}
  {2008})}\BibitemShut {NoStop}%
\bibitem [{\citenamefont {Tarasenko}(2009)}]{Tarasenko09}%
  \BibitemOpen
  \bibfield  {author} {\bibinfo {author} {\bibfnamefont {S.~A.}\ \bibnamefont
  {Tarasenko}},\ }\href {http://dx.doi.org/10.1103/PhysRevB.80.165317}
  {\bibfield  {journal} {\bibinfo  {journal} {Phys. Rev. B}\ }\textbf {\bibinfo
  {volume} {80}},\ \bibinfo {pages} {165317} (\bibinfo {year}
  {2009})}\BibitemShut {NoStop}%
\bibitem [{\citenamefont {V\"olkl}\ \emph {et~al.}(2011)\citenamefont
  {V\"olkl}, \citenamefont {Griesbeck}, \citenamefont {Tarasenko},
  \citenamefont {Schuh}, \citenamefont {Wegscheider}, \citenamefont
  {Sch\"uller},\ and\ \citenamefont {Korn}}]{Volkl2011}%
  \BibitemOpen
  \bibfield  {author} {\bibinfo {author} {\bibfnamefont {R.}~\bibnamefont
  {V\"olkl}}, \bibinfo {author} {\bibfnamefont {M.}~\bibnamefont {Griesbeck}},
  \bibinfo {author} {\bibfnamefont {S.~A.}\ \bibnamefont {Tarasenko}}, \bibinfo
  {author} {\bibfnamefont {D.}~\bibnamefont {Schuh}}, \bibinfo {author}
  {\bibfnamefont {W.}~\bibnamefont {Wegscheider}}, \bibinfo {author}
  {\bibfnamefont {C.}~\bibnamefont {Sch\"uller}}, \ and\ \bibinfo {author}
  {\bibfnamefont {T.}~\bibnamefont {Korn}},\ }\href {\doibase
  10.1103/PhysRevB.83.241306} {\bibfield  {journal} {\bibinfo  {journal} {Phys.
  Rev. B}\ }\textbf {\bibinfo {volume} {83}},\ \bibinfo {pages} {241306}
  (\bibinfo {year} {2011})}\BibitemShut {NoStop}%
\bibitem [{\citenamefont {Poshakinskiy}\ and\ \citenamefont
  {Tarasenko}(2013)}]{Poshakinskiy2013}%
  \BibitemOpen
  \bibfield  {author} {\bibinfo {author} {\bibfnamefont {A.~V.}\ \bibnamefont
  {Poshakinskiy}}\ and\ \bibinfo {author} {\bibfnamefont {S.~A.}\ \bibnamefont
  {Tarasenko}},\ }\href {\doibase 10.1103/PhysRevB.87.235301} {\bibfield
  {journal} {\bibinfo  {journal} {Phys. Rev. B}\ }\textbf {\bibinfo {volume}
  {87}},\ \bibinfo {pages} {235301} (\bibinfo {year} {2013})}\BibitemShut
  {NoStop}%
\bibitem [{\citenamefont {Pikus}\ and\ \citenamefont
  {Titkov}(1984)}]{Pikus_OO}%
  \BibitemOpen
  \bibfield  {author} {\bibinfo {author} {\bibfnamefont {G.~E.}\ \bibnamefont
  {Pikus}}\ and\ \bibinfo {author} {\bibfnamefont {A.}~\bibnamefont {Titkov}},\
  }\href@noop {} {\emph {\bibinfo {title} {Optical Orientation}}},\ edited by\
  \bibinfo {editor} {\bibfnamefont {F.}~\bibnamefont {Mayer}}\ and\ \bibinfo
  {editor} {\bibfnamefont {B.}~\bibnamefont {Zakharchenya}}\ (\bibinfo
  {publisher} {North Holland, Amsterdam},\ \bibinfo {year} {1984})\BibitemShut
  {NoStop}%
\bibitem [{\citenamefont {Pikus}\ \emph {et~al.}(1988)\citenamefont {Pikus},
  \citenamefont {Maruschak},\ and\ \citenamefont {Titkov}}]{Pikus88}%
  \BibitemOpen
  \bibfield  {author} {\bibinfo {author} {\bibfnamefont {G.~E.}\ \bibnamefont
  {Pikus}}, \bibinfo {author} {\bibfnamefont {V.~A.}\ \bibnamefont
  {Maruschak}}, \ and\ \bibinfo {author} {\bibfnamefont {A.~N.}\ \bibnamefont
  {Titkov}},\ }\href@noop {} {\bibfield  {journal} {\bibinfo  {journal} {Sov.
  Phys. Semicond.}\ }\textbf {\bibinfo {volume} {22}},\ \bibinfo {pages} {115}
  (\bibinfo {year} {1988})},\ \translation{[Fiz. Tekh. Poluprovdn. {\bf 22},
  185 (1988)}\BibitemShut {NoStop}%
\bibitem [{\citenamefont {Matsuda}\ and\ \citenamefont
  {Yoh}(2011)}]{Matsuda11}%
  \BibitemOpen
  \bibfield  {author} {\bibinfo {author} {\bibfnamefont {T.}~\bibnamefont
  {Matsuda}}\ and\ \bibinfo {author} {\bibfnamefont {K.}~\bibnamefont {Yoh}},\
  }\href {http://www.sciencedirect.com/science/article/pii/S0022024810011929}
  {\bibfield  {journal} {\bibinfo  {journal} {Journal of Crystal Growth}\
  }\textbf {\bibinfo {volume} {323}},\ \bibinfo {pages} {52} (\bibinfo {year}
  {2011})}\BibitemShut {NoStop}%
\bibitem [{\citenamefont {Nestoklon}\ \emph {et~al.}(2012)\citenamefont
  {Nestoklon}, \citenamefont {Tarasenko}, \citenamefont {Jancu},\ and\
  \citenamefont {Voisin}}]{Nestoklon12}%
  \BibitemOpen
  \bibfield  {author} {\bibinfo {author} {\bibfnamefont {M.~O.}\ \bibnamefont
  {Nestoklon}}, \bibinfo {author} {\bibfnamefont {S.~A.}\ \bibnamefont
  {Tarasenko}}, \bibinfo {author} {\bibfnamefont {J.-M.}\ \bibnamefont
  {Jancu}}, \ and\ \bibinfo {author} {\bibfnamefont {P.}~\bibnamefont
  {Voisin}},\ }\href {\doibase 10.1103/PhysRevB.85.205307} {\bibfield
  {journal} {\bibinfo  {journal} {Phys. Rev. B}\ }\textbf {\bibinfo {volume}
  {85}},\ \bibinfo {pages} {205307} (\bibinfo {year} {2012})}\BibitemShut
  {NoStop}%
\bibitem [{\citenamefont {Chantis}\ \emph {et~al.}(2008)\citenamefont
  {Chantis}, \citenamefont {Cardona}, \citenamefont {Christensen},
  \citenamefont {Smith}, \citenamefont {van Schilfgaarde}, \citenamefont
  {Kotani}, \citenamefont {Svane},\ and\ \citenamefont {Albers}}]{Chantis08}%
  \BibitemOpen
  \bibfield  {author} {\bibinfo {author} {\bibfnamefont {A.~N.}\ \bibnamefont
  {Chantis}}, \bibinfo {author} {\bibfnamefont {M.}~\bibnamefont {Cardona}},
  \bibinfo {author} {\bibfnamefont {N.~E.}\ \bibnamefont {Christensen}},
  \bibinfo {author} {\bibfnamefont {D.~L.}\ \bibnamefont {Smith}}, \bibinfo
  {author} {\bibfnamefont {M.}~\bibnamefont {van Schilfgaarde}}, \bibinfo
  {author} {\bibfnamefont {T.}~\bibnamefont {Kotani}}, \bibinfo {author}
  {\bibfnamefont {A.}~\bibnamefont {Svane}}, \ and\ \bibinfo {author}
  {\bibfnamefont {R.~C.}\ \bibnamefont {Albers}},\ }\href {\doibase
  10.1103/PhysRevB.78.075208} {\bibfield  {journal} {\bibinfo  {journal} {Phys.
  Rev. B}\ }\textbf {\bibinfo {volume} {78}},\ \bibinfo {pages} {075208}
  (\bibinfo {year} {2008})}\BibitemShut {NoStop}%
\bibitem [{\citenamefont {Beck}\ \emph {et~al.}(2006)\citenamefont {Beck},
  \citenamefont {Metzner}, \citenamefont {Malzer},\ and\ \citenamefont
  {D{\"o}hler}}]{Beck06}%
  \BibitemOpen
  \bibfield  {author} {\bibinfo {author} {\bibfnamefont {M.}~\bibnamefont
  {Beck}}, \bibinfo {author} {\bibfnamefont {C.}~\bibnamefont {Metzner}},
  \bibinfo {author} {\bibfnamefont {S.}~\bibnamefont {Malzer}}, \ and\ \bibinfo
  {author} {\bibfnamefont {G.~H.}\ \bibnamefont {D{\"o}hler}},\ }\href
  {\doibase 10.1209/epl/i2006-10151-4} {\bibfield  {journal} {\bibinfo
  {journal} {Europhys. Lett.}\ }\textbf {\bibinfo {volume} {75}},\ \bibinfo
  {pages} {597} (\bibinfo {year} {2006})}\BibitemShut {NoStop}%
\bibitem [{\citenamefont {Jancu}\ \emph {et~al.}(1998)\citenamefont {Jancu},
  \citenamefont {Scholz}, \citenamefont {Beltram},\ and\ \citenamefont
  {Bassani}}]{Jancu98}%
  \BibitemOpen
  \bibfield  {author} {\bibinfo {author} {\bibfnamefont {J.-M.}\ \bibnamefont
  {Jancu}}, \bibinfo {author} {\bibfnamefont {R.}~\bibnamefont {Scholz}},
  \bibinfo {author} {\bibfnamefont {F.}~\bibnamefont {Beltram}}, \ and\
  \bibinfo {author} {\bibfnamefont {F.}~\bibnamefont {Bassani}},\ }\href
  {\doibase 10.1103/PhysRevB.57.6493} {\bibfield  {journal} {\bibinfo
  {journal} {Phys. Rev. B}\ }\textbf {\bibinfo {volume} {57}},\ \bibinfo
  {pages} {6493} (\bibinfo {year} {1998})}\BibitemShut {NoStop}%
\bibitem [{\citenamefont {Keating}(1966)}]{Keating66}%
  \BibitemOpen
  \bibfield  {author} {\bibinfo {author} {\bibfnamefont {P.~N.}\ \bibnamefont
  {Keating}},\ }\href {\doibase 10.1103/PhysRev.145.637} {\bibfield  {journal}
  {\bibinfo  {journal} {Phys. Rev.}\ }\textbf {\bibinfo {volume} {145}},\
  \bibinfo {pages} {637} (\bibinfo {year} {1966})}\BibitemShut {NoStop}%
\bibitem [{\citenamefont {Steiger}\ \emph {et~al.}(2011)\citenamefont
  {Steiger}, \citenamefont {Salmani-Jelodar}, \citenamefont {Areshkin},
  \citenamefont {Paul}, \citenamefont {Kubis}, \citenamefont {Povolotskyi},
  \citenamefont {Park},\ and\ \citenamefont {Klimeck}}]{Steiger11}%
  \BibitemOpen
  \bibfield  {author} {\bibinfo {author} {\bibfnamefont {S.}~\bibnamefont
  {Steiger}}, \bibinfo {author} {\bibfnamefont {M.}~\bibnamefont
  {Salmani-Jelodar}}, \bibinfo {author} {\bibfnamefont {D.}~\bibnamefont
  {Areshkin}}, \bibinfo {author} {\bibfnamefont {A.}~\bibnamefont {Paul}},
  \bibinfo {author} {\bibfnamefont {T.}~\bibnamefont {Kubis}}, \bibinfo
  {author} {\bibfnamefont {M.}~\bibnamefont {Povolotskyi}}, \bibinfo {author}
  {\bibfnamefont {H.-H.}\ \bibnamefont {Park}}, \ and\ \bibinfo {author}
  {\bibfnamefont {G.}~\bibnamefont {Klimeck}},\ }\href {\doibase
  10.1103/PhysRevB.84.155204} {\bibfield  {journal} {\bibinfo  {journal} {Phys.
  Rev. B}\ }\textbf {\bibinfo {volume} {84}},\ \bibinfo {pages} {155204}
  (\bibinfo {year} {2011})}\BibitemShut {NoStop}%
\bibitem [{\citenamefont {Pryor}\ \emph {et~al.}(1998)\citenamefont {Pryor},
  \citenamefont {Kim}, \citenamefont {Wang}, \citenamefont {Williamson},\ and\
  \citenamefont {Zunger}}]{Pryor98}%
  \BibitemOpen
  \bibfield  {author} {\bibinfo {author} {\bibfnamefont {C.}~\bibnamefont
  {Pryor}}, \bibinfo {author} {\bibfnamefont {J.}~\bibnamefont {Kim}}, \bibinfo
  {author} {\bibfnamefont {L.~W.}\ \bibnamefont {Wang}}, \bibinfo {author}
  {\bibfnamefont {A.~J.}\ \bibnamefont {Williamson}}, \ and\ \bibinfo {author}
  {\bibfnamefont {A.}~\bibnamefont {Zunger}},\ }\href {\doibase
  http://dx.doi.org/10.1063/1.366631} {\bibfield  {journal} {\bibinfo
  {journal} {Journal of Applied Physics}\ }\textbf {\bibinfo {volume} {83}},\
  \bibinfo {pages} {2548} (\bibinfo {year} {1998})}\BibitemShut {NoStop}%
\bibitem [{\citenamefont {Jancu}\ and\ \citenamefont {Voisin}(2007)}]{Jancu07}%
  \BibitemOpen
  \bibfield  {author} {\bibinfo {author} {\bibfnamefont {J.-M.}\ \bibnamefont
  {Jancu}}\ and\ \bibinfo {author} {\bibfnamefont {P.}~\bibnamefont {Voisin}},\
  }\href {\doibase 10.1103/PhysRevB.76.115202} {\bibfield  {journal} {\bibinfo
  {journal} {Phys. Rev. B}\ }\textbf {\bibinfo {volume} {76}},\ \bibinfo
  {pages} {115202} (\bibinfo {year} {2007})}\BibitemShut {NoStop}%
\bibitem [{\citenamefont {Boykin}\ \emph {et~al.}(2010)\citenamefont {Boykin},
  \citenamefont {Luisier}, \citenamefont {Salmani-Jelodar},\ and\ \citenamefont
  {Klimeck}}]{Boykin10}%
  \BibitemOpen
  \bibfield  {author} {\bibinfo {author} {\bibfnamefont {T.~B.}\ \bibnamefont
  {Boykin}}, \bibinfo {author} {\bibfnamefont {M.}~\bibnamefont {Luisier}},
  \bibinfo {author} {\bibfnamefont {M.}~\bibnamefont {Salmani-Jelodar}}, \ and\
  \bibinfo {author} {\bibfnamefont {G.}~\bibnamefont {Klimeck}},\ }\href
  {\doibase 10.1103/PhysRevB.81.125202} {\bibfield  {journal} {\bibinfo
  {journal} {Phys. Rev. B}\ }\textbf {\bibinfo {volume} {81}},\ \bibinfo
  {pages} {125202} (\bibinfo {year} {2010})}\BibitemShut {NoStop}%
\bibitem [{\citenamefont {Niquet}\ \emph {et~al.}(2009)\citenamefont {Niquet},
  \citenamefont {Rideau}, \citenamefont {Tavernier}, \citenamefont {Jaouen},\
  and\ \citenamefont {Blase}}]{Niquet09}%
  \BibitemOpen
  \bibfield  {author} {\bibinfo {author} {\bibfnamefont {Y.~M.}\ \bibnamefont
  {Niquet}}, \bibinfo {author} {\bibfnamefont {D.}~\bibnamefont {Rideau}},
  \bibinfo {author} {\bibfnamefont {C.}~\bibnamefont {Tavernier}}, \bibinfo
  {author} {\bibfnamefont {H.}~\bibnamefont {Jaouen}}, \ and\ \bibinfo {author}
  {\bibfnamefont {X.}~\bibnamefont {Blase}},\ }\href {\doibase
  10.1103/PhysRevB.79.245201} {\bibfield  {journal} {\bibinfo  {journal} {Phys.
  Rev. B}\ }\textbf {\bibinfo {volume} {79}},\ \bibinfo {pages} {245201}
  (\bibinfo {year} {2009})}\BibitemShut {NoStop}%
\bibitem [{\citenamefont {Zieli\ifmmode~\acute{n}\else
  \'{n}\fi{}ski}(2012)}]{Zielinski12}%
  \BibitemOpen
  \bibfield  {author} {\bibinfo {author} {\bibfnamefont {M.}~\bibnamefont
  {Zieli\ifmmode~\acute{n}\else \'{n}\fi{}ski}},\ }\href {\doibase
  10.1103/PhysRevB.86.115424} {\bibfield  {journal} {\bibinfo  {journal} {Phys.
  Rev. B}\ }\textbf {\bibinfo {volume} {86}},\ \bibinfo {pages} {115424}
  (\bibinfo {year} {2012})}\BibitemShut {NoStop}%
\bibitem [{\citenamefont {Raouafi}\ \emph {et~al.}(2016)\citenamefont
  {Raouafi}, \citenamefont {Benchamekh}, \citenamefont {Nestoklon},
  \citenamefont {Jancu},\ and\ \citenamefont {Voisin}}]{Raouafi16}%
  \BibitemOpen
  \bibfield  {author} {\bibinfo {author} {\bibfnamefont {F.}~\bibnamefont
  {Raouafi}}, \bibinfo {author} {\bibfnamefont {R.}~\bibnamefont {Benchamekh}},
  \bibinfo {author} {\bibfnamefont {M.~O.}\ \bibnamefont {Nestoklon}}, \bibinfo
  {author} {\bibfnamefont {J.-M.}\ \bibnamefont {Jancu}}, \ and\ \bibinfo
  {author} {\bibfnamefont {P.}~\bibnamefont {Voisin}},\ }\href
  {http://stacks.iop.org/0953-8984/28/i=4/a=045001} {\bibfield  {journal}
  {\bibinfo  {journal} {Journal of Physics: Condensed Matter}\ }\textbf
  {\bibinfo {volume} {28}},\ \bibinfo {pages} {045001} (\bibinfo {year}
  {2016})}\BibitemShut {NoStop}%
\bibitem [{\citenamefont {Vurgaftman}\ \emph {et~al.}(2001)\citenamefont
  {Vurgaftman}, \citenamefont {Meyer},\ and\ \citenamefont
  {Ram-Mohan}}]{Meyer}%
  \BibitemOpen
  \bibfield  {author} {\bibinfo {author} {\bibfnamefont {I.}~\bibnamefont
  {Vurgaftman}}, \bibinfo {author} {\bibfnamefont {J.~R.}\ \bibnamefont
  {Meyer}}, \ and\ \bibinfo {author} {\bibfnamefont {L.~R.}\ \bibnamefont
  {Ram-Mohan}},\ }\href {\doibase http://dx.doi.org/10.1063/1.1368156}
  {\bibfield  {journal} {\bibinfo  {journal} {Journal of Applied Physics}\
  }\textbf {\bibinfo {volume} {89}},\ \bibinfo {pages} {5815} (\bibinfo {year}
  {2001})}\BibitemShut {NoStop}%
\bibitem [{\citenamefont {Shim}\ and\ \citenamefont {Rabitz}(1998)}]{Shim98}%
  \BibitemOpen
  \bibfield  {author} {\bibinfo {author} {\bibfnamefont {K.}~\bibnamefont
  {Shim}}\ and\ \bibinfo {author} {\bibfnamefont {H.}~\bibnamefont {Rabitz}},\
  }\href {\doibase 10.1103/PhysRevB.57.12874} {\bibfield  {journal} {\bibinfo
  {journal} {Phys. Rev. B}\ }\textbf {\bibinfo {volume} {57}},\ \bibinfo
  {pages} {12874} (\bibinfo {year} {1998})}\BibitemShut {NoStop}%
\bibitem [{\citenamefont {Nestoklon}\ \emph {et~al.}(2016)\citenamefont
  {Nestoklon}, \citenamefont {Benchamekh},\ and\ \citenamefont
  {Voisin}}]{Nestoklon16}%
  \BibitemOpen
  \bibfield  {author} {\bibinfo {author} {\bibfnamefont {M.}~\bibnamefont
  {Nestoklon}}, \bibinfo {author} {\bibfnamefont {R.}~\bibnamefont
  {Benchamekh}}, \ and\ \bibinfo {author} {\bibfnamefont {P.}~\bibnamefont
  {Voisin}},\ }\href {http://arxiv.org/abs/1603.03227} {\bibfield  {journal}
  {\bibinfo  {journal} {Journal of Physics: Condensed Matter}\ }\textbf
  {\bibinfo {volume} {??}},\ \bibinfo {pages} {?????} (\bibinfo {year}
  {2016})}\BibitemShut {NoStop}%
\bibitem [{Note1()}]{Note1}%
  \BibitemOpen
  \bibinfo {note} {The coordinate system is changed with respect to the one
  used in Ref.~\cite {Nestoklon12} which results in the opposite sign of the
  Dresselhaus constant.}\BibitemShut {Stop}%
\bibitem [{\citenamefont {Boykin}(1998)}]{Boykin98}%
  \BibitemOpen
  \bibfield  {author} {\bibinfo {author} {\bibfnamefont {T.~B.}\ \bibnamefont
  {Boykin}},\ }\href {\doibase 10.1103/PhysRevB.57.1620} {\bibfield  {journal}
  {\bibinfo  {journal} {Phys. Rev. B}\ }\textbf {\bibinfo {volume} {57}},\
  \bibinfo {pages} {1620} (\bibinfo {year} {1998})}\BibitemShut {NoStop}%
\end{thebibliography}%

\end{document}